\documentclass[conference]{IEEEtran}
\IEEEoverridecommandlockouts
\usepackage{cite}
\usepackage{amsmath,amssymb,amsfonts}
\usepackage{algorithmic,algorithm}
\usepackage{graphicx}
\usepackage{textcomp}
\usepackage{xcolor}
\usepackage{subfigure}
\def\BibTeX{{\rm B\kern-.05em{\sc i\kern-.025em b}\kern-.08em
		T\kern-.1667em\lower.7ex\hbox{E}\kern-.125emX}}
\begin{document}
	
	\title{
		Multi-Agent Graph Reinforcement Learning based On-Demand Wireless Energy Transfer in Multi-UAV-aided IoT Network
	}
	\author{\IEEEauthorblockN{Ze Yu Zhao$^1$, Yueling Che$^2$, Sheng Luo$^3$, Kaishun Wu$^4$, and Victor C. M. Leung$^5$}
		\IEEEauthorblockA{College of Computer Science and Software Engineering, Shenzhen University, China\\ Email: 2100271103$^1$@email.szu.edu.cn, \{yuelingche$^2$, sluo$^3$, wu$^4$, vleung$^5$\}@szu.edu.cn}
	} \maketitle

\begin{abstract}

This paper proposes a new on-demand wireless energy transfer (WET)  scheme of multiple unmanned aerial vehicles (UAVs). Unlike the existing studies that simply pursuing the total or the minimum harvested energy maximization at the Internet of Things (IoT) devices, where the IoT devices' own energy requirements are barely considered, we propose a new  metric called the hungry-level of energy (HoE), which reflects the time-varying energy demand of each IoT device based on the energy gap between its required energy and the harvested energy from the UAVs. 
With the purpose to minimize the overall HoE of the IoT devices whose energy requirements are not satisfied, we optimally determine  all the UAVs' trajectories and WET decisions over time, under the practical mobility and energy constraints of the UAVs. Although the proposed problem is of high complexity to solve, by excavating the UAVs' self-attentions for their collaborative WET, we propose the multi-agent graph reinforcement learning (MAGRL) based approach. Through the offline training of the MAGRL model, where the global training at the central controller guides the   local training at each UAV agent, each UAV then distributively determines its trajectory and WET based on the well-trained local neural networks. Simulation results show that the proposed MAGRL-based approach outperforms various benchmarks for meeting the IoT devices' energy requirements.
 
\end{abstract}

\newtheorem{definition}{\underline{Definition}}[section]
\newtheorem{fact}{Fact}
\newtheorem{assumption}{Assumption}
\newtheorem{theorem}{\underline{Theorem}}[section]
\newtheorem{lemma}{\underline{Lemma}}[section]
\newtheorem{corollary}{\underline{Corollary}}[section]
\newtheorem{proposition}{\underline{Proposition}}[section]
\newtheorem{example}{\underline{Example}}[section]
\newtheorem{remark}{\underline{Remark}}[section]
\newtheorem{observation}{\underline{Observation}}[section]
\newcommand{\mv}[1]{\mbox{\boldmath{$ #1 $}}}

\section{Introduction}
 
The technology of radio frequency (RF) based wireless energy transfer (WET) has been recognized as a promising approach to support energy-sustainable Internet of Things (IoT) networks\cite{ref.1}. Conventionally, ground infrastructure (e.g., dedicated energy transmitters or base stations) are utilized to charge the low-power IoT devices. However, due to the generally low end-to-end wireless energy transmission efficiency, to assure non-zero harvested energy at the IoT devices, the effective transmission distance from the ground infrastructure to each IoT device is restricted (to , e.g., 10 meters \cite{ref.3}). 

By exploiting the UAVs' flexible mobility to effectively shorten the transmission distances between the UAVs and the IoT devices, the UAV-aided WET has attracted a great deal of attentions. For example, the UAV's dynamic wireless energy and information transmission  scheme in the presence of primary users was investigated in \cite{Yue Ling Che}. The UAV's hovering position to maximize the minimum harvested energy at the IoT devices was studied in \cite{ref.7}. The UAV's trajectory and WET were jointly optimized for maximizing the total harvested energy at the IoT devices in \cite{ref.6}.

The above studies only considered the WET by a single UAV, which is usually difficult to serve a large-scale network due to its limited on-board battery energy. 
By tackling with the more complicated joint design of the multiple UAVs' trajectories and wireless energy transmissions, the multi-UAV aided WET for maximizing the total harvested energy at the IoT devices has been studied in \cite{ref.19} and \cite{ref.20}, where the Lagrange multiplier method and the fly-and-hover based trajectory design were utilized, respectively. The joint design of multi-UAV-aided wireless data and energy transmissions has also been investigated in \cite{ref.8} and \cite{ref.9}, where the deep reinforcement learning (DRL) is applied to adapt the multiple UAVs' transmissions to the dynamic environment.

However, it came to our notice that the existing UAV-aided WET schemes may sacrifice the energy demands of some IoT devices for achieving higher global benefits. For example, with the purpose to maximize the total harvested energy at all the IoT devices, the UAVs with limited on-board energy may choose to transmit energy to the closely-located IoT devices more often, but barely fly to serve the distantly-located IoT devices to save energy; and with the purpose to maximize the minimum harvested energy at all the IoT devices, the IoT devices with high energy demands may not be able to harvest sufficient energy, for achieving equal energy harvesting at each IoT device. As a result, both designs may deviate from the IoT devices' own energy demands. 

To cater to the IoT devices' energy requirements, we propose a new metric called hungry-level of energy (HoE), which reflects the time-varying energy desirability of each IoT device based on the energy gap between its required energy and the harvested energy from the UAVs. Moreover, to explore the UAVs' potential collaborations, such that they can automatically determine their joint or separate WET depending on the IoT devices' HoE, we employ the UAVs' self-attentions based on the graph-based representation. Finally, we propose the novel multi-agent graph reinforcement learning (MAGRL) based approach to minimize the overall HoE of the IoT devices whose energy requirements are not satisfied.

\begin{figure}
	\setlength{\abovecaptionskip}{-0.0in}
	\centering
	\DeclareGraphicsExtensions{.eps,.mps,.pdf,.jpg,.png}
	\DeclareGraphicsRule{*}{eps}{*}{}
	\includegraphics[angle=0, width=0.48 \textwidth]{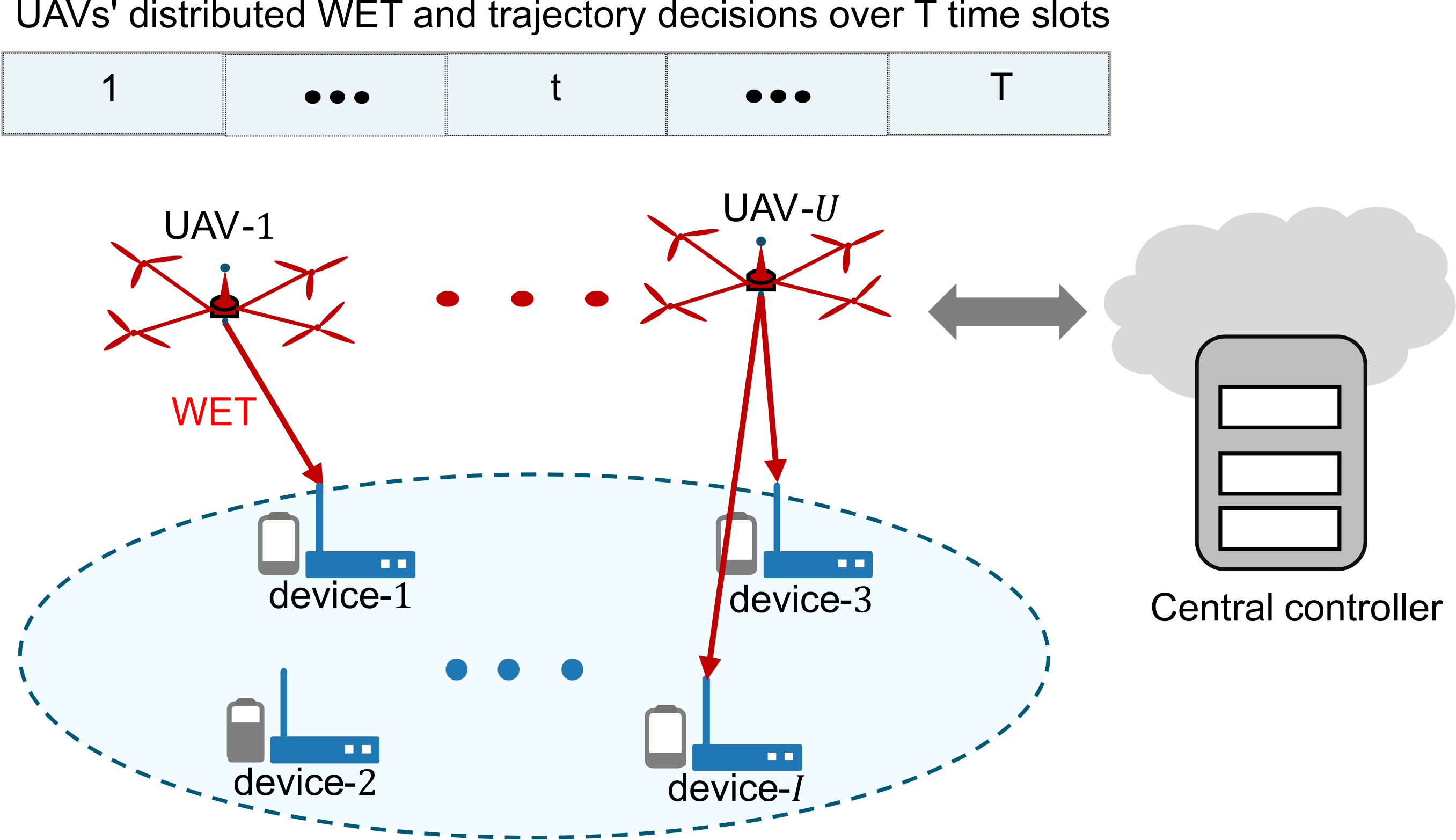}
	\caption{Multi-UAV assisted WET in an IoT system.}
	\label{fig: system_model}
	\vspace{-0.25in}
\end{figure}

The main contributions are summarized as follows:
\begin{itemize}
\item \emph{HoE-based Multi-UAV WET Modeling and Novel Problem Formulation}: Section \ref{section: system model} newly proposes the metric of HoE to guide the multi-UAV-aided WET for satisfying the different energy demands of the IoT devices. Based on each IoT device's non-linear energy harvesting model and each UAV's velocity-determined energy consumption model, the battery energy management at both the UAVs and the IoT devices are properly modeled. By optimally determining all the UAVs' trajectories and WET decisions over time, the novel HoE minimization problem is formulated under the UAVs' practical mobility and energy constraints.

 \item \emph{MAGRL-based Approach for Distributed and Collaborative Multi-UAV WET}: Sections \ref{section MDP and Global} and \ref{section: MAGR} propose the MAGRL-based approach to solve the complicated HoE minimization problem, where the UAVs' self-attentions are excavated for their collaborative WET. Through the offline training of the proposed MAGRL model, where the central controller leads the global training to guide the local training at each UAV agent, each UAV then distributively determines its trajectory and WET decision based on the well-trained policy neural networks.

 \item \emph{Extensive Simulation Results for Performance Evaluation}: Section \ref{section: perform} conducts extensive simulations to verify the validity of the proposed HoE metric for guiding the UAVs' on-demand WET, by comparing with various benchmarks. The UAVs' 
 collaborative WET under the proposed MAGRL-based approach is also illustrated.
\end{itemize}

\section{System Model and the Problem Formulation}\label{section: system model}

As shown in Fig. \ref{fig: system_model}, we consider that $U$ UAVs with $U\geq 2$ act as the airborne wireless energy transmitters to charge in total of $I$ IoT devices with low power consumptions on ground. Each UAV flies at a fixed altitude of $h_{fix}$ meters (m).
Denote the sets of the UAVs and the IoT devices as $\mathbb{U}=\{1,2,...,U\}$ and $\mathbb{I}=\{1,2,...,I\}$, respectively. The UAVs' task period of WET is divided into $T$ time slots with $T\geq 1$, where the slot length is $\vartheta$ seconds (s). The set of the time slots is denoted as $\mathbb{T}=\{1,2,...,T\}$. The coordinate of UAV-$u$ in slot $t$ is represented as $q_u [t]=\{x_u [t],y_u [t],h_{fix}\}$, $\forall u\in \mathbb{U}$, $\forall t\in \mathbb{T}$. Since the slot length $\vartheta$ is usually very small, $q_u[t]$ is assumed to be unchanged in each slot, but may change over different slots. The coordinate of device-$i$ with $i\in \mathbb{I}$ on ground is denoted as $q_i=\{x_i,y_i,0\}$. Let $d_i^u [t]=\left\|q_u [t]-q_i \right\|$ denote the distance between UAV-$u$ and device-$i$ in slot $t$. Let $C_u[t]\in \{0,1\}$ denote UAV-$u$'s WET decision in slot $t$, where UAV-$u$ broadcasts energy in slot $t$ if $C_u(t)=1$, or keeps silent, otherwise.

According to \cite{ref.22}, denote $P_{Los,i}^u [t]=\frac{1}{1+a \exp(-b\beta_i^u [t]+ab) }$ as the line-of-sight (LoS) probability of the air-to-ground (AtG) channel from UAV-$u$ to device-$i$ in slot $t$, where $\beta_i^u [t]=\sin^{-1} \left( \frac{h_{fix}}{d_i^u [t]}\right) $ is the elevation angles from UAV-$u$ to device-$i$ in slot $t$, and the constants $a$ and $b$ are the environment-related parameters. We then obtain the non-line-of-sight (NLoS) probability as $P_{NLoS,i}^u [t] = 1-P_{Los,i}^u [t]$. As a result, the average AtG channel power gain from UAV-$u$ to device-$i$ is
\begin{equation}
	\label{equ: channel Gain}
	G_i^u[t]= P_{LoS,i}^u[t]G_0d_i^u[t]^{\!-\!\alpha _L}\!+\! P_{NLoS,i}^u[t]G_0d_i^u[t]^{\!-\!\alpha _N}\!,\!
\end{equation}
where $G_0$ is the average channel gain at a reference distance of $1$ m, and $\alpha_L$ and $\alpha_N$ are the channel path-loss exponents for LoS and NoS links, respectively.

\subsection{UAV Energy Consumption Model}

Each UAV's energy consumption is mainly caused by propulsion and WET. According to \cite{ref.23}, in each slot $t$,  UAV-$u$'s propulsion power consumption is determined by its velocity $V_u[t]=\frac{1}{\vartheta}\left\|q_u[t+1]-q_u[t]\right\|$ as follows:
	\begin{align}
		\label{equ: UAV Propulsion Power}
		P_{pro}(V_u[t])&=P_a\left(1+\frac{3V_u[t]^2}{V_{tip}^2}\right )+ \frac{1}{2}f_0\rho e_1AV_u[t]^3  \nonumber  \\ 
		&+P_b\left (\sqrt{1+\frac{V_u[t]^4}{4e_0^4}}-\frac{V_u[t]^2}{2e_0^2}\right )^{\frac{1}{2} },
	\end{align}
where the constants $P_a $, $P_b$, $ V_{tip}$, $e_0$, $e_1$, $f_0$ and $\rho$ are the UAV's mechanical-related parameters. Hence, the propulsion energy consumption of UAV-$u$ in slot $t$ is obtained as $P_{pro}(V_u [t])\vartheta$. Denote $P_u$ as each UAV-$u$'s transmit power for WET. The energy consumption of UAV-$u$'s WET in slot $t$ is thus $C_u [t] P_u \vartheta$. Denote $B_u[t]$ as the battery level of UAV-$u$ at the beginning of slot $t$, which is updated as
\begin{align}
	\label{equ: UAV Energy}
	B_u[t] \!  = \! \max \!  \left( B_u[t \! -\! 1]\!-\! P_{pro}(V_u[t \!-\!1]) \vartheta \!-\! C_u[t \!-\!1]P_u\vartheta, 0\right).
\end{align}
We assume that UAV-$u$ is fully charged in the initial with $B_u [0] = B_u^{max}$, where $B_u^{max}$ is UAV-$u$'s battery capacity.

\subsection{Energy Harvesting at IoT devices}

Each IoT device is installed with a rectenna and an energy harvester. According to \cite{ref.24}, the energy harvester transforms the received RF power $p\geq0$ at the rectenna into the direct-circuit (DC) power $\mathcal{F}(p)$ as follows:
\begin{equation}
	\label{equ: Energy Harvesting Conversion}
	\mathcal{F}(p)=\begin{cases}
		0,& p\in [0, P_{sen}), \\
		f(p),& p\in[P_{sen}, P_{sat}), \\
		f(P_{sat}),& p \in [P_{sat}, +\infty],
	\end{cases}
\end{equation}
where $P_{sen}$ and $P_{sat}$ with $0<P_{sen}<P_{sat}$ are the sensitivity power and the saturation power at the energy harvester, respectively, and $f(\cdot)$ is a non-linear power transform function that can be easily obtained through the curve fitting technique \cite{ref.24}. From (\ref{equ: Energy Harvesting Conversion}), no DC power is harvested if $p$ is below $P_{sen}$ and the harvested DC power keeps unchanged if $p \ge P_{sat}$. The value of $P_{sen}$ is usually high (with, e.g., -10 dBm \cite{ref.PowerCast}) in practice. Hence, the transmission distance from the UAV to the IoT device needs to be sufficiently short to assure effective WET with non-zero harvested energy. For each device-$i$, since its harvested energy from multiple UAVs in each slot can be accumulated, based on (\ref{equ: channel Gain}) and (\ref{equ: Energy Harvesting Conversion}), the harvested energy at device-$i$ in slot $t$ is obtained as
\begin{equation}
	\label{equ: harvested energy}
	E_i^{har} [t]=\mathcal{F} \left (\sum_{u=1}^{U}P_uC_u[t]G_i^u[t] \right) \vartheta.
\end{equation}
From (\ref{equ: harvested energy}), device-$i$ can harvest more energy if more UAVs are located nearby and transmit energy to it jointly.  Denote $B_i [t]$ as the battery level of device-$i$ at the beginning of slot $t$, which is updated as
\begin{equation}
		\label{equ:  IoT Energy}
		B_i[t] = \min \left( B_i[t-1] + E_i^{har}[t-1], B_i^{max} \right),
\end{equation}
where $B_i^{max}$ is the battery capacity of device-$i$ and the initial battery energy $B_i [0] \geq 0 $ is given. Denote $B^{thr}$ as the required battery energy level at each IoT device, where $B_i^{max}\ge B^{thr}>B_i[0]>0$ holds in general. We say an IoT device is \emph{energy satisfied} if its accumulated battery energy reaches the required $B^{thr}$ before or at the last slot $T$. Hence, $\mathbb{I}_l [t]=\{i|B_i [t]+E_i^{har}[t]<B^{thr}\}$ is the set of energy-unsatisfied IoT devices at the end of slot $t$. After the UAVs' WET for $T$ slots, the set of energy-unsatisfied IoT devices is obtained as $\mathbb{I}_l[T]$.

\subsection{Hunger-Level of Energy at IoT devices}

The HoE is defined to measure the time-varying energy demand of each IoT device. Denote $H_i[t]$ as the HoE of device-$i$ at the beginning of slot $t$. We define the time variation of $H_i[t]$ as
	\begin{equation}
		\label{equ: HoE}
		H_i[t]=\begin{cases}
			\max (H_i[t \!-\! 1]\!-\! 1, \! 1),& \! \textrm{if~} E_i^{har}[t \!-\! 1]\! \ge \! E^{exp} \! \textrm{~and~} \\& \! B_i[t]< B^{thr}, \\
			H_i[t \! - \! 1]\!+\! 1, & \! \textrm{if~} E_i^{har}[t \! - \! 1]\!<\! E^{exp} \textrm{~and~} \\& B_i[t] \!<\! B^{thr}, \\
			0, & \! \textrm{if~} B_i[t] \! \ge \! B^{thr},  
		\end{cases}
	\end{equation}
where $E^{exp}\! \triangleq \! \frac{B^{thr}}{T}$ denotes the average amount of energy that device-$i$ expects to harvest in each slot for reaching $B^{thr}$ after $T$ slots. If device-$i$'s harvested energy $E_{i}^{har}[t-1]$ in slot ($t-1$) reaches the expected $E^{exp}$, but the resultant battery energy $B_i[t]$ at the beginning of slot $t$ is still lower than the required $B^{thr}$, $H_i[t]$ is reduced by $1$ at the beginning of slot $t$, where the minimum allowable HoE when $B_i[t]<B^{thr}$ is set to $1$. If $E_{i}^{har}[t-1]$  is lower than the expected $E^{exp}$ and the resultant $B_i[t]$ is also lower than the required $B^{thr}$, $H_i[t]$ is increased by $1$ at the beginning of slot $t$. Moreover, if device-$i$'s required energy is satisfied with $B_i[t]\geq B^{thr}$ at the beginning of slot $t$, $H_i[t]$ becomes $0$. The overall HoE of all the energy-unsatisfied IoT devices over $T$ slots is then obtained as
\begin{equation}
	\label{equ: total HoE}
	H_{total}=
	\sum_{i\in \mathbb{I}_l[T]} \sum_{t=1}^{T} H_i[t].
\end{equation}
It is easy to find that $H_{total}$ is $0$, if $\mathbb{I}_l[T]$ is empty.

\subsection{Problem Formulation}
By optimally determining all the UAVs' WET decisions $\boldsymbol{C}=\{C_u[t] \}$ and trajectories $\boldsymbol{Q}=\{ q_u[t] \}$ over $T$ slots, we minimize the overall HoE $H_{total}$ in (\ref{equ: total HoE}) under the UAVs' practical mobility and energy constraints as follows:
\begin{align} 
	\textrm{(P1)}:~\min_{\boldsymbol{Q},\boldsymbol{C}}& 
	 \sum_{i\in \mathbb{I}_l[T]} \sum_{t=1}^{T} H_i[t], \nonumber \\ 
	\mathrm{s.t.}  
	& (\ref{equ: UAV Energy}),~(\ref{equ:  IoT Energy}),~(\ref{equ: HoE}) \nonumber \\ 
	& \left \| q_u[t] \!-\! q_u[t \!-\!1] \right \| \!\leq\! V_u^{max}\vartheta , \forall u \!\in\! \mathbb{U}, \forall t \!\in \!\mathbb{T},  \label{equ: constraint a} \\
	& C_u[t]\in\{0,1\}, \forall u \in \mathbb{U}, \forall t \in \mathbb{T},  \label{equ: constraint b} \\
	&B_u[T]\ge B_u^{min}, \forall u \in \mathbb{U}, \label{equ: constraint c} \\
	&d_u^{u^{'}}[t]\ge d_{min}, \forall u,u^{'} \in \mathbb{U}, u\neq u^{'} ,\forall t \in \mathbb{T}, \label{equ: constraint d}\\
	&x_u[t]\! \in \! \left[0,\! W_{max}\!\right],\! y_u[t]\! \in \! \left[ 0,\! L_{max} \right],  \!  \forall u \! \in \! \mathbb{U}, \forall t \! \in \! \mathbb{T}. \label{equ: constraint e}
\end{align}
The constraint in (\ref{equ: constraint a}) ensures that the velocity of UAV-$u$ does not exceed its maximal allowable velocity $V_u^{max}$. The constraint in (\ref{equ: constraint b}) gives each UAV's binary WET decision. The constraint in (\ref{equ: constraint c}) ensures that each UAV's remained energy at the end of slot $T$ is no less than the minimum required energy $B_u^{min}$ for a safe return after the WET task. The constraint in (\ref{equ: constraint d}) guarantees a safe distance between any two UAVs in each slot to avoid collisions. The constraint in (\ref{equ: constraint e}) confines each UAV's horizontal moving space within an area of length $L_{max}$ and width $W_{max}$.

Problem (P1) is a mixed-integer programming problem. It is also noticed that with the goal to minimize the overall HoE, the multiple UAVs need to be efficiently organized, by either jointly transmitting energy to the same set of IoT devices  that are closely located, or separately serving different sets of IoT devices that are distantly located. Hence, all the UAVs' trajectories and WET decisions are naturally coupled with each other over time. Moreover, as constrained by (\ref{equ: constraint c}), each UAV must use the limited battery energy wisely for reducing the IoT devices' HoE. Therefore, problem (P1) is generally difficult to solve efficiently by using the traditional optimization methods.

\section{MDP Modeling and Global Graph Design} \label{section MDP and Global}

Considering the above complicated and coupled relations in problem (P1) among multiple UAVs, a multi-agent DRL approach is leveraged in this paper. As shown in Fig. \ref{fig: system_model}, each UAV acts as an agent and reports its environment states to the central controller (e.g., a base station or a satellite); By using the global environment information of all the UAVs, the central controller's training output also guides each UAV's local training. Although the training is centralized, after the training process, each UAV distributively determines its own WET decision and trajectory based on its local policy. Moreover, to explore the potential collaboration of all the UAVs for efficient WET, we take the global UAV information as a graph and introduce the similarity matrix \cite{ref.26} and the self-attention block \cite{ref.27} to operate the graph-based global information at the central controller \cite{ref.28}. By doing so, a new MAGRL-based approach is proposed to solve problem (P1). In this section, we model the Markov decision processes (MDP) at each UAV, and then introduce the UAVs' graph-based representations at the central controller. The MAGRL-based solution will be specified in Section \ref{section: MAGR}.

\subsection{MDP Modeling}
According to problem (P1), by letting each UAV act as an agent, we model the MDP for each of the $U$ agents\cite{ref.30}. For each agent, define the MDP as a set of states $\mathbb{S}$, a set of actions $\mathbb{A}$, and a set of rewards $\mathbb{R}$. The state set $\mathbb{S}$ embraces all the possible environment configurations at each UAV, including the UAV's own location, the HoE of all the IoT devices, the battery levels of all the IoT devices \footnote{It is assumed that all the IoT devices share their HoE and battery levels with the UAVs via a common channel.}, and the UAV's own battery level. The action set $\mathbb{A}$ provides the action space of each UAV's decision on its trajectory and WET. 
For any given state $s_u [t]\in \mathbb{S}$ for UAV-$u$ at the beginning of slot $t$, UAV-$u$ applies the policy $\pi_u:s_u [t] \to a_u [t]$ to select the action $a_u [t]\in \mathbb{A}$, and then gets the corresponding reward $r_u [t]\in \mathbb{R}$ at the end of slot $t$.

Specifically, the state in slot $t$ is defined as $s_u[t]=\left\{ \! x_u[t] \! ,\! y_u[t]\!,\! H_1[t]\!,...,\! H_I[t]\! ,\! B_1[t]\!,...,\! B_I[t]\!,\! B_u[t]\!\right\}$, which contains in total of $M=2I+3$ elements.
Denoting $\varphi_u[t]$ as UAV-$u$'s horizontal rotation angle in slot $t$, UAV-$u$'s horizontal location $\left(x_u[t], y_u[t]\right)$ in problem (P1) is determined if $\varphi_u[t]$ and $V_u[t]$ are obtained. Thus UAV-$u$'s MDP action is defined as $a_{u}[t]= \left\{ V_u[t], \varphi_u[t], C_u[t] \right\}$, where $C_u[t]=0$ if the policy network output is negative or $C_u[t]=1$, otherwise. The reward function is proposed as
\begin{equation}
	\label{equ: local reward}
	r_{u}[t]=\xi_0r_{u,0}[t]-\xi_1r_{u,1}[t],
\end{equation}
where $r_{u,0} [t],r_{u,1} [t]$ are the reward and penalty that UAV-$u$ receives in slot $t$, respectively, and $\xi_0, \xi_1\in(0,1)$ are the corresponding weights. Specifically, letting $w_i^u \triangleq \frac{\mathcal{F}(P_uC_u[t]G_i^u[t])\vartheta}{E_i^{har}[t]}$ denote the ratio of the DC energy that device-$i$ harvestes from UAV-$u$ to that from all the UAVs, we use $N_u[t]=\frac{w_u[t]}{\sum_{u\in\mathbb{U}}w_u[t]}$ with $w_u[t]=\sum_{i\in\mathbb{I}}w_i^u[t]$ to represent UAV-$u$'s  effective WET weight among all the UAVs in slot $t$. The IoT devices can harvest higher amounts of energy from the UAV with a higher $N_u[t]$, and vice versa. We then propose to use the following $r_{u,0}[t]$:  
\begin{align}   
		r_{u,0}[t]=\frac{N_u[t]\sum_{i\in\mathbb{I}_l[t]}(B_i[t+1]-B_i[t])\cdot H_i[t]}
		{1+|\mathbb{I}_l[t]|\sum_{i\in\mathbb{I}_l[t]}H_i[t]} \nonumber \\ + \xi_2(B_u[t+1]-B_u^{min}). \label{equ: local reward a}
\end{align}
From (\ref{equ: local reward a}), while the UAVs prefer to perform WET more frequently to reduce the IoT devices' HoE, they also need to use their battery energy carefully to assure the constraint in (\ref{equ: constraint c}). Hence, the reward in  (\ref{equ: local reward a}) contains two items. The first item is UAV-$u$'s reward for charging IoT devices, where the numerator is the product of UAV-$u$'s weight $N_u[t]$ and all the IoT devices harvested energy biased by their HoE, and the denominator is the product of the HoE summation over all the energy-unsatisfied IoT devices and the set size $|\mathbb{I}_l[t]|$, which plus $1$ to prevent the denominator from being $0$. It is easy to find that, the more energy the high-HoE IoT devices can harvest, the higher value the first item achieves. The second item is UAV-$u$'s battery energy gap between $B_u[t+1] $ and  $B_u^{min}$ in constraint (\ref{equ: constraint c}) after taking action $a_u[t]$ in slot $t$, where a balance parameter $\xi_2$ is multiplied. The penalty in (\ref{equ: local reward}) is designed as 
\begin{equation}  
	r_{u,1}[t]= \sum_{u=1}^{U}\sum_{j=0}^{1}PEN_u^j, \label{equ: local reward b} 
\end{equation}
where $PEN_u^0=1$ (or $PEN_u^1=1$) if the constraint in (\ref{equ: constraint d}) (or (\ref{equ: constraint e})) is not satisfied, or $PEN_u^0=0$ (or $PEN_u^1=0$), otherwise.

\subsection{Graph Representation of UAVs} \label{subsection: graph representation}

By receiving each UAV's state information, the central controller obtains the global information. Let $\boldsymbol{o}[t]=\{s_1[t],...,s_U[t]\}$, $\boldsymbol{a}[t]=\{a_1[t],...,a_U[t]\}$ and $\boldsymbol{r}[t]=\{r_1[t],...,r_U[t]\}$ denote the global observations, actions, and rewards, respectively. To explore the potential connections among the UAVs to improve the overall WET performance as well as to avoid collisions, the central controller uses a graph to represent all the UAVs, by treating each UAV as a node in the graph.
According to \cite{ref.26}, we use the following similarity matrix among the UAVs to represent the strength of their connections,
\begin{equation}  
\boldsymbol{Z}[t]=
\begin{pmatrix}
	z_{11}[t],&...   &,z_{1U}[t] \\
	...&  &... \\
	z_{U1}[t],&...  &,z_{UU}[t]
\end{pmatrix}_{U\times U} , \label{equ: similarity matrix}
\end{equation}
where the element $z_{uu^{'}} [t]=\exp \left( -\frac{\left\| q_u[t]-q_{u^{'}}[t] \right\|^2}{2\varrho^2} \right)$, $\forall u, u^{'}\in \mathbb{U}, u\neq u^{'}$, is the Gaussian distance between UAV-$u$ and UAV-$u^{'}$, $\varrho^2$ is a constant, and the element $z_{uu} [t]=\sum_{u^{'}\in \mathbb{U}, u^{'}\neq u}z_{uu^{'}}[t]$ on the diagonal is used as the degree of UAV-$u$.
Specifically, from \cite{ref.27}, to obtain the global feature matrix $\tilde{\boldsymbol{o}}$, the central controller first generates the attention matrix $\boldsymbol{W}_{att}$ and value matrix $\boldsymbol{W}_v^{'}$ as follows: 
\begin{align}
	&\boldsymbol{W}_{att} = softmax\left( \! \frac{1}{\sqrt{M}}\left( \boldsymbol{o} \! \times \! \boldsymbol{W}_q \right) \! \times \! \left( \boldsymbol{o} \! \times \! \boldsymbol{W}_k \right)^T \! \cdot \! \boldsymbol{Z} \! \right), \nonumber \\   &\boldsymbol{W}_v^{'}=\boldsymbol{o}\times \boldsymbol{W}_v,  \label{equ: self-attention matrix}
\end{align}
where the symbol $\times$ denotes the matrix multiplication, $(\cdot)^T$ is the matrix transposition. For any matrix $\boldsymbol{Z}\in \mathbf{R}^{U\times U}$, the $softmax(\cdot)$ function transforms the element $z_{uu^{'}}$ into $\frac{\exp \left(z_{uu^{'}}\right)}{\sum_{u^{'}\in \mathbb{U}} \exp \left(z_{uu^{'}}\right)}$, $\forall u,u^{'}\in \mathbb{U}$. Then, the global feature matrix $\tilde{\boldsymbol{o}}$ is obtained as $\tilde{\boldsymbol{o}}=\boldsymbol{W}_{att} \times \boldsymbol{W}_v^{'}+ \boldsymbol{W}_v^{'}$, which is used as the new observation matrix for the central controller.

\begin{figure*}
	\setlength{\abovecaptionskip}{-0.0in}
	\centering
	\DeclareGraphicsExtensions{.eps,.mps,.pdf,.jpg,.png}
	\DeclareGraphicsRule{*}{eps}{*}{}
	\includegraphics[angle=0, width=0.95 \textwidth]{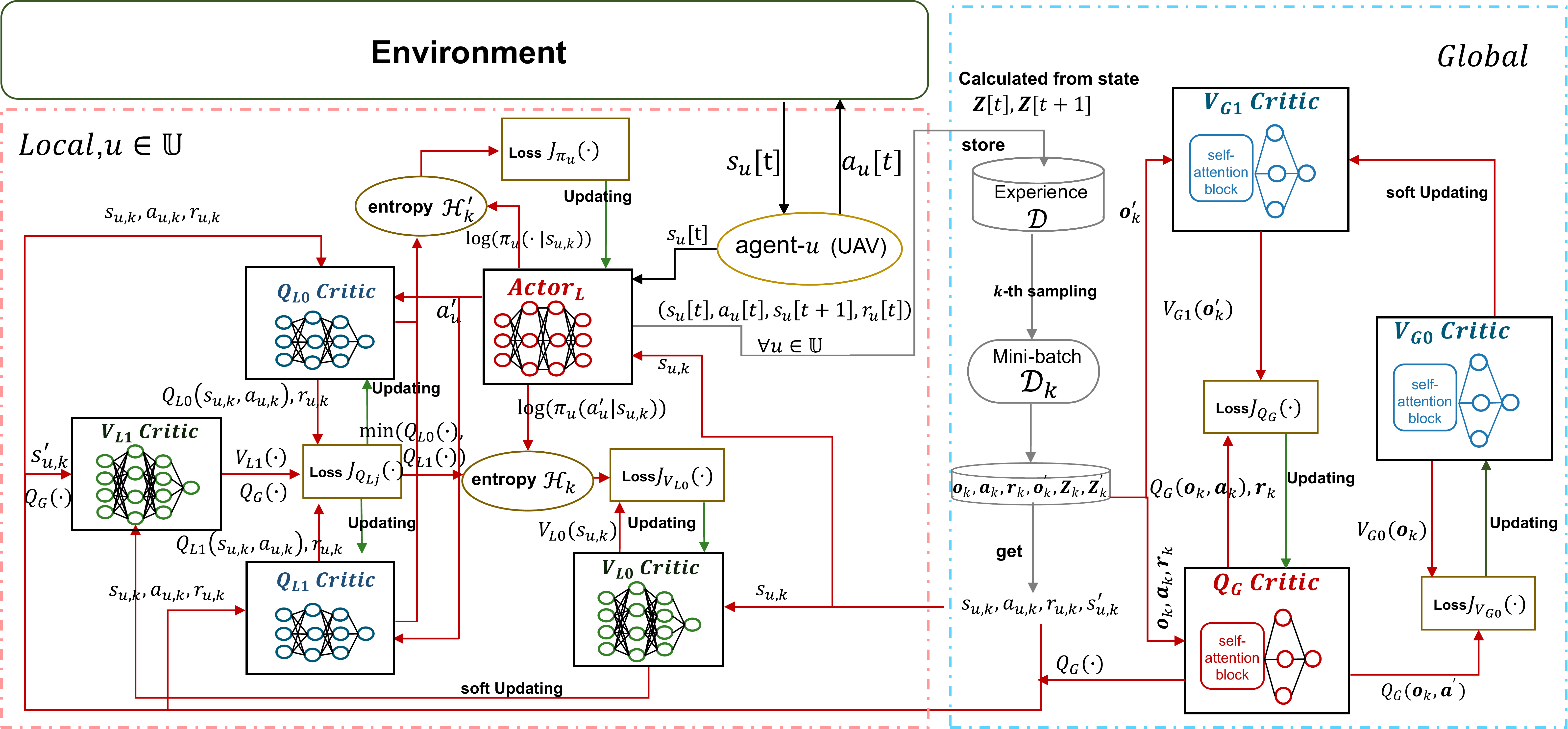}
	\caption{MAGRL framework. }
	\label{fig: Training}
	\vspace{-0.1in}
\end{figure*}

\section{MAGRL-based Solution}\label{section: MAGR}

\subsection{MAGRL Training Flow}
As shown in Fig. \ref{fig: Training}, the MAGRL framework includes two parts, where one is the local training at each of the UAV, and the other is the global training at the central controller. In training stage, a tuple $\left( \boldsymbol{o}[t],\boldsymbol{a}[t],\boldsymbol{r}[t],\boldsymbol{o}[t \!+\! 1],\boldsymbol{Z}[t],\boldsymbol{Z}[t \!+\! 1] \right)$ is stored in the experience replay buffer $\mathcal{D}$, and all the neural networks for both local and global training apply the stochastic gradient descent (SGD) algorithm to update their parameters.

\textbf{Local Training}: For each UAV's local training, we apply the SAC algorithm proposed in \cite{ref.29} to enhance the exploration of the environment for all agents. As shown in the left side of Fig. \ref{fig: Training}, for any UAV-$u$, there are five neural networks employed for its local training, which are the policy network $Actor_L$, the local Q-networks $Q_{L0}~Critic$ and $Q_{L1}~Critic$, and the local V-networks $V_{L0}~Critic$ and $V_{L1}~Critic$, with the corresponding network parameters denoted by $\theta^{\pi_u}$, $\eta_{0,u}$, $\eta_{1,u}$, $\phi_{0,u}$ and $\phi_{1,u}$, respectively. The policy network is trained for the policy function $\pi_u (\cdot)$ that maps UAV-$u$'s state $s_u[t]$ to its action $a_u[t]$, the two local Q-networks are trained for the local state action functions $Q_{L0} (\cdot)$ and $Q_{L1}(\cdot)$, and the two local V-networks are trained for the local state functions $V_{L0} (\cdot)$ and $V_{L1}(\cdot)$. The information entropy $\mathcal{H}(\cdot)$ is used to enhance the agent's exploration of the environment \cite{ref.29}. The goal of the local training is to obtain the optimal policy $\pi_u^{*}=\arg\max_{\pi_u}\sum_{t\in \mathbb{T}}\mathbb{E}\left[ r_u[t]+\alpha_u \mathcal{H}\left( \pi_u(\cdot | s_u[t]) \right) \right]$, where the temperature coefficient $\alpha_u$ is the weight of the information entropy, and $\mathbb{E}[\cdot]$ is the expectation operation over all the possible actions. The performance of UAV-$u$ when taking action $a_u[t]$ in state $s_u [t]$ is evaluated by the local Q-networks $Q_{L0}~Critic$ and $Q_{L1}~Critic$, with $Q_{Lj}(s_u[t], a_u[t])=r_u[t]+\gamma \mathbb{E}[V_{L1}(s_u[t \!+\!1])]$, $j\in \{0,1\}$. The performance of UAV-$u$ in state $s_u [t]$ is evaluated by the local V-network $V_{L1}~Critic$, with $V_{L1}(s_u[t])=\mathbb{E}_{a_u^{'}\sim \pi_u} [ Q_{min}(s_u[t],a_u[t])- \alpha_u\log ( \pi_u (a_u^{'}|s_u[t]))]$, where $a_u^{'} \sim \pi_u$ denotes the action taken from policy $\pi_u$, and $Q_{min} (\cdot)=\min(Q_{L0} (\cdot),Q _{L1} (\cdot))$.

The parameters of all the five neural networks are updated based on the corresponding loss functions. Specifically, after receiving the $k$-th local experience $(s_{u,k},a_{u,k},r_{u,k},s_{u,k}^{'})$ from the mini-batch $\mathcal{D}_k$ of central controller for the local training, UAV-$u$ uses the loss function
\begin{align}  
		J_{V_{L0}}(\phi_{0,u})\!=\!\mathbb{E}\left[\! \frac{1}{2}\left(\!
		V_{L0}(s_{u,k};\! \phi_{0,u})- Q_{exp}
		\right)^2 \! \right]  \label{equ: Loss local V} 
\end{align}
to update $\phi_{0,u}$ for the local V-network $V_{L0}~Critic$ in negative the direction of the gradient $\widehat{\nabla}_{\phi_{0,u}}J_{V_{L0}}(\phi_{0,u})$, where $Q_{exp} \!\triangleq \!\mathbb{E}_{a_u^{'}\sim\! \pi_u} \left[ \! Q_{min}(s_{u,k},a_u^{'}; \!\eta_{j,u})\!-\!\alpha_u \mathcal{H}_k \! \right]$ is the expected entropy-added local minimum Q-value and $\mathcal{H}_k \!\triangleq\! \log \left(\! \pi_u (a_u^{'}| s_{u,k};\! \theta^{\pi_u})\! \right)$ is the entropy. For the parameter $\phi_{1,u}$ of $V_{L1} ~Critic$ network, we perform a soft update via $\phi_{1,u} \gets \tau \phi_{1,u} + (1-\tau) \phi_{0,u}$, $\tau \in [0,1)$. For the update of the parameters $\eta_{0,u}$ and $\eta_{1,u}$ for $Q_{L0}~Critic$ and $Q_{L1}~Critic$ networks, respectively, they are also computed in the negative direction of the corresponding loss function's gradient with
\begin{align}  
		J_{Q_{Lj}}(\eta_{j,u})\!=\!\mathbb{E} \left[ \frac{1}{2} \left( Q_{Lj} (s_{u,k},a_{u,k}; \eta_{j,u}) \!-\! y_{Lj} \right)^2 \right], \label{equ: Loss local Q} 
\end{align}
where $j\in\{0,1\}$ and by using $Q_G(\cdot)$ to denote the global Q-value used to guide the training of the two local Q-networks, we define $y_{Lj}\! \triangleq \!\epsilon \!\left(\! r_{u,k}\! +\! \gamma \mathbb{E}\! \left[\!V_{L1}(s_{u,k}^{'};\! \phi_{1,u})\! \right] \!\right) \!+ \!(1\!-\epsilon) \mathbb{E}\left[Q_G(\cdot) \right]$ with $\epsilon\in(0,1]$.
Similarly, according to the loss function for the policy network
\begin{equation}  
	J_{\pi_u}(\theta^{\pi_u}) = \mathbb{E}_{\varepsilon \sim \mathcal{N}} \left[ \alpha_u \mathcal{H}_k^{'}\! - \! Q_{min}\left( s_{u,k}, f_{\theta^{\pi_u}}; \eta_{j,u} \right) \right],  \label{equ: Loss policy} 
\end{equation}
the network parameter $\theta^{\pi_u}$ is updated in the negative direction of the gradient $\widehat{\nabla}_{\theta^{\pi_u}}J_{\pi_u}(\theta^{\pi_u})$, where the information entropy $\mathcal{H}_k^{'} \triangleq \log (\pi_u \left( f_{\theta^{\pi_u}}(s_{u,k}; \varepsilon) | s_{u,k} \right))$ is calculated from the noise-added action $f_{\theta^{\pi_u}}(s_{u,k}; \varepsilon)$, and $\varepsilon$ is the noise sampled from a fixed distribution $\mathcal{N}$. Based on \cite{ref.29}, adding noise to the action prevents the network from overfitting and ensures the stable network training. We use the loss function
	\begin{equation}  
		J_{\alpha_u}(\alpha_u)=\mathbb{E}_{a_u^{'}\sim \pi_u} \left[  
		-\alpha_u \log \left( \pi_u (a_u^{'}| s_{u,k}; \theta^{\pi_u}) \right) -\alpha_u \tilde{\mathcal{H}}
		\right]  \label{equ: Loss entropy} 
	\end{equation}
to update the temperature coefficient $\alpha_u$ in the negative direction of the gradient $\widehat{\nabla}_{\alpha_u}J_{\alpha_u}(\alpha_u)$ with $\tilde{\mathcal{H}} \triangleq | s_u[t] |$.

\textbf{Global Training}: The global training is designed for the central controller which consists of three neural networks, which are the global Q-network $Q_G~Critic$ and the two global V-networks $V_{G0}~Critic$ and $V_{G1}~Critic$, with the corresponding network parameters denoted as $\eta_G$, $\phi_{G0}$ and $\phi_{G1}$, respectively. The global Q-network is trained for the global state action function $Q_G (\cdot)$ and the two global V-networks are trained for the global state functions $V_{G0} (\cdot)$ and $V_{G1}(\cdot)$. As shown in the right side of Fig. \ref{fig: Training}, each neural network contains the self-attention block, which extracts the global feature matrix $\tilde{\boldsymbol{o}}$ to obtain the connections among UAVs as specified in Section \ref{subsection: graph representation}. The goal of the central training is to find the optimal global Q-value
$Q_G^{*}(\boldsymbol{o}[t],\boldsymbol{a}[t])=\boldsymbol{r}+\gamma\mathbb{E}[V_{G1}(\boldsymbol{s}[t])]$.

The parameters of the three networks are also updated based on the corresponding loss functions. Specifically, after receiving the $k$-th global experience ($\boldsymbol{o}_k, \boldsymbol{a}_k, \boldsymbol{r}_k, \boldsymbol{o}^{'}_k,\boldsymbol{Z}_k,\boldsymbol{Z}_k^{'}$) from the mini-batch $\mathcal{D}_k$, the central controller uses the loss function
\vspace{-0.3em}
\begin{equation}  
	J_{V_{G0}}(\phi_{G0})=\mathbb{E}\left[  \!
	\frac{1}{2} \! \left( V_{G0}(\boldsymbol{o}_k;\phi_{G0}) \! - \! \mathbb{E}_{\boldsymbol{a}^{'}\sim \pi_{u}} \! Q_G(\boldsymbol{o}_k,\! \boldsymbol{a}^{'}; \! \eta_G) \! \right)^2 \!
	\right]  \label{equ: Loss global V} 
\end{equation}
to update $\phi_{G0}$ for the global V-network $V_{G0}~Critic$ in the negative direction of the gradient $\widehat{\nabla}_{\phi_{G0}}J_{V_{G0}}(\phi_{G0})$. For the parameter $\phi_{G1}$ of $V_{G1~Critic}$ network, we perform a soft update via $\phi_{G1} \gets \tau \phi_{G1} + (1-\tau) \phi_{G0}$, $\tau \in [0,1)$. For the update of the parameter $\eta_G$ for $Q_G~Critic$ network, it is also computed in the negative direction of the corresponding loss function's gradient, where the loss function is given as
\begin{equation}  
	J_{\! Q_G \!}(\! \eta_G \!)\!=\!\mathbb{E}\! \left[ \!\frac{1}{2}\! \left(\!
	Q_G(\boldsymbol{o}_k,\!\boldsymbol{a}_k;\! \eta_G) \!-\! \left( \!\boldsymbol{r}_k \!+\! \gamma \mathbb{E}\left[\! V_{G1}(\boldsymbol{o}_k^{'}; \!\phi_{G1}) \!\right] \!\right)\!
	\right)^2\! \right]\!.  \label{equ: Loss global Q} 
\end{equation}

\subsection{MAGRL-Based Algorithm}

Based on the above framework, we propose the MAGRL-based algorithm to solve problem \textrm{(P1)}. The MAGRL-based algorithm is specified in \textbf{Algorithm 1}.

\begin{figure*}[htbp]
	\centering 
	\subfigure[Reward variation.]{ \label{fig: MAGR performance a}
		\begin{minipage}{7.5cm}
			\centering
			\includegraphics[width=7.5cm]{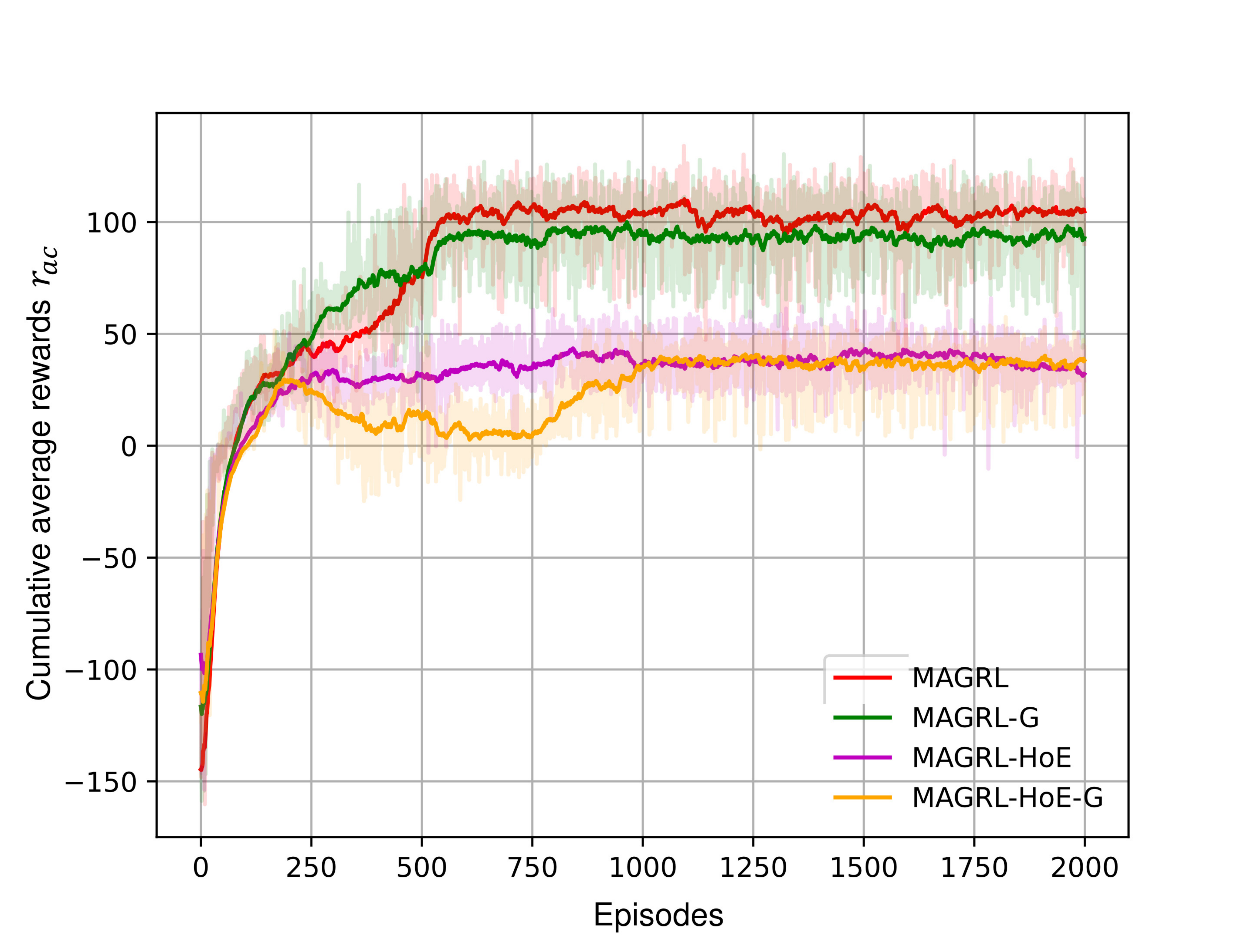}
		\end{minipage}
	}
	\subfigure[$H_{total}$ variation.]{ \label{fig: MAGR performance b}
		\begin{minipage}{7.5cm}
			\centering
			\includegraphics[width=7.5cm]{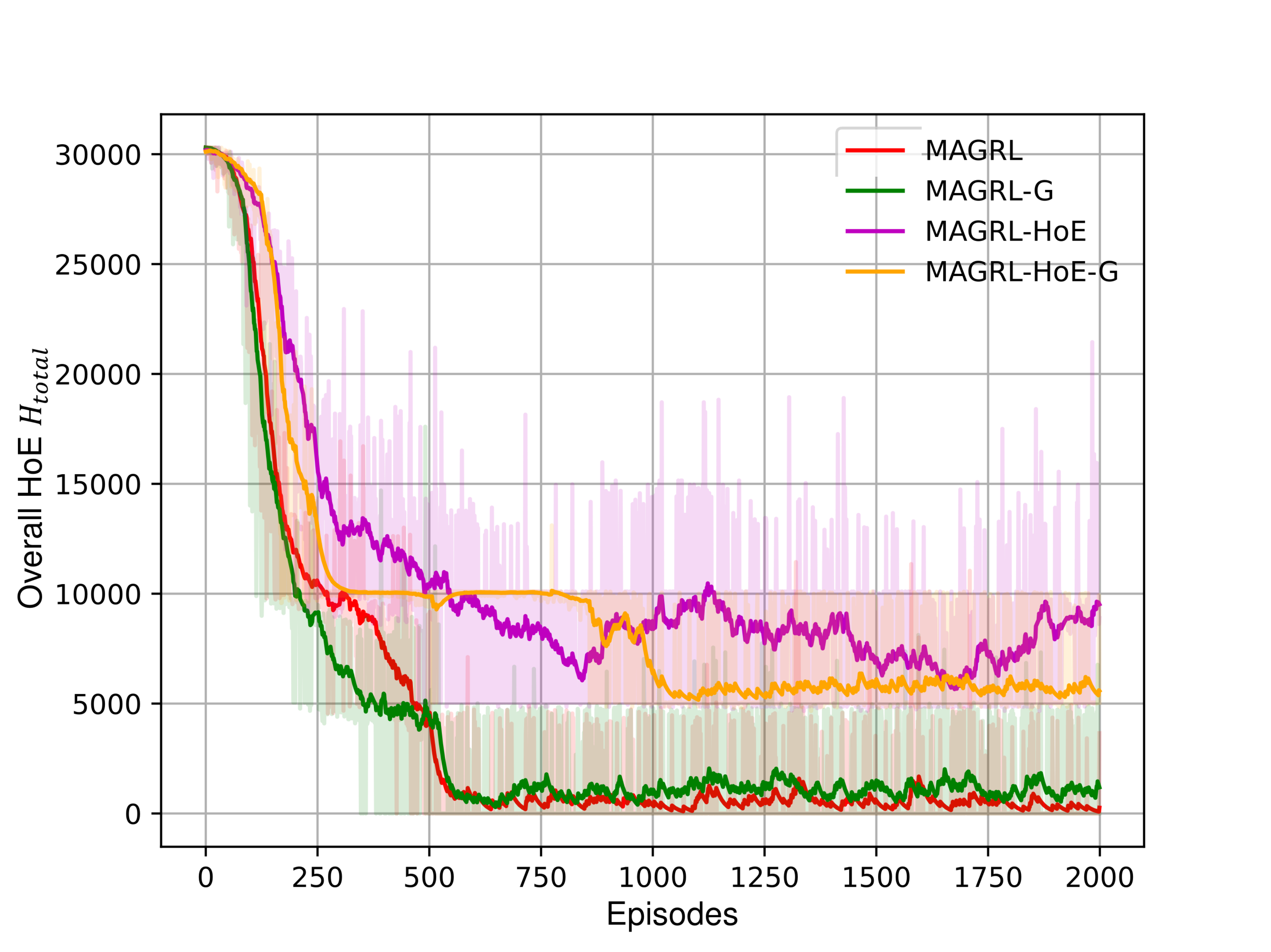}
		\end{minipage}
	}
	\vspace{-0.1in}
	\caption{Comparison of MAGR-method to baseline methods.}
\end{figure*}

\begin{table}[h]
	\centering
	\caption{ Simulation parameters}\label{table: parameters set}
	\vspace{-1em}
	\scriptsize
	\setlength{\tabcolsep}{1  mm}{
		\begin{tabular}{|c|c||c|c|}
			\hline
			\textbf{Parameter}   & \textbf{Value}   & \textbf{Parameter}   & \textbf{Value}      \\ \hline
			$h_{fix}$       &  $5$ m  &  $\sigma^2,\varrho^2$  & $-90$ dBm, 100 \\ \hline
			$P_u$  & $1$ W  &  $P_{sen}, P_{sat}$  & $-10$,$7$ dBm \\ \hline
			$\alpha_L, \alpha_N$       &  $3,5$  &  $B^{thr}$  & $10$ mW$\cdot$s \\ \hline
			$B_u^{min}$, $B_u^{max}$      &  $20000$,$140000$ W$\cdot$s &  $d_{min}$   &   $5$ m   \\ \hline
			$\gamma,\epsilon,\tau$  &  $0.985$, $0.8$, $0.999$         &  $a,b$ for $P_{LoS}$   & $12.08$, $0.11$   \\ \hline
			$\xi_0,\xi_1,\xi_2$  & $0.25$, $1$, $0.00001$ &   Size of $\mathcal{D}$ &  $2^{17}$   \\ \hline
			size of $\mathcal{D}_k$   & $128$   &  $\alpha_u$'s learning rate  &  $0.0002$ \\ \hline
	\end{tabular}}
	\vspace{-2em}
\end{table}

\section{Simulation Results} \label{section: perform}

To evaluate the performance of our proposed MAGRL method, we conduct simulations based on python-3.9.12 and pytorch-1.12.1. 
Unless specified otherwise, in all the simulations, the starting horizontal position of each UAV is selected randomly in the the considered area, and each IoT device's initial battery energy is set randomly in the range between $2$ mW$\cdot$s and $5$ mW$\cdot$s, respectively. The UAV's propulsion model parameters are set as in \cite{ref.24}. Each neural network for the local training has $4$ layers and the number of neurons in the hidden layer is $256$. Each neural network in global training contains a self-atteion block and $2$ fully connected layers. Learning rate is set to $0.0002$ for all neural networks except for the neural network $Actor_L$, which has a learning rate of $0.0003$. Other parameters can be found in Table \ref{table: parameters set}.

\begin{algorithm} \label{algorithm: 1}
	\small
	\caption{MAGRL-based solution}
	\begin{algorithmic}[1]
		\STATE {Initialize replay buffer $\mathcal{D}$, learning rate $\lambda$, discount factor $\gamma$, soft update weight $\tau$ and temperature factor $\alpha_u$, $u\in\mathbb{U}$. Initialize the parameters of local's all five networks and global three networks}
		\FOR{Episode $\gets1,...,EPS$}
		\STATE{Initialize the location and energy of all UAVs and IoT devices;}
		\STATE{Initialize the observatin $\boldsymbol{o}[0]$ and similarity matrix $B[0]$;}
		\FOR{$t\gets 1,..., T$}
		\STATE{\textbf{get action} $a_{u}[t]=\pi_{u}(s_{u}[t]|\theta^{\pi_{u}})$, $u\in\mathbb{U}$}
		\STATE{\textbf{execute action} $a_{u}[t]=\left[V_u[t],\omega_u[t],C_u[t]\right]$, $\forall u\in\mathbb{U}$. We can get $\boldsymbol{o}[t+1]$, $\boldsymbol{r}[t]$ and $\boldsymbol{Z}[t+1]$;}
		\STATE{\textbf{store} $\left(\boldsymbol{o}[t], \boldsymbol{a}[t], \boldsymbol{r}[t], \boldsymbol{o}[t+1], \boldsymbol{Z}[t], \boldsymbol{Z}[t+1] \right)$ into experience replay buffer $\mathcal{D}$;}
		\IF{$|\mathcal{D}| \ge$mini-batch of size $\triangle$}
		\FOR{$u\gets 1,...,U$}
		\STATE{$\phi_{0,u} \gets \phi_{0,u}-\lambda  \widehat{\nabla}_{\phi_{0,u}}J_{V_{L0}}(\phi_{0,u})$, update $\phi_{0,u}$;}
		\STATE{$\eta_{j,u}\gets\eta_{j,u}-\lambda \widehat{\nabla}_{\eta_{j,u}}J_{Q_{Lj}}(\eta_{j,u}),j\in \left\{0,1\right\}$;}
		\STATE{$\theta^{\pi_{u}}\gets\theta^{\pi_{u}}-\lambda\widehat{\nabla}_{\theta^{\pi_{u}}}J_{\pi_u}\left( \theta^{\pi_{u}}\right)$;}
		\STATE{$\alpha_u\gets \alpha_u -\lambda\widehat{\nabla}_{\alpha_u}J_{\alpha_u}(\alpha_u)$, update tempture factor based on(\ref{equ: Loss entropy});}
		\STATE{$\phi_{1,u}\gets\tau\phi_{1,u}+(1-\tau)\phi_{0,u}$, soft update;}
		\ENDFOR
		\STATE{$\phi_{G0} \gets \phi_{G0}-\lambda  \widehat{\nabla}_{\phi_{G0}}J_{V_{G0}}(\phi_{G0})$, update $\phi_{G0}$;}
		\STATE{$\eta_{G}\gets\eta_{G}-\lambda \widehat{\nabla}_{\eta_{G}}J_{Q_{G}}(\eta_{G})$;}
		\STATE{$\phi_{G1}\gets\tau\phi_{G1}+(1-\tau)\phi_{G0}$, soft update;}
		\ENDIF
		\STATE{$\boldsymbol{o}[t]\gets \boldsymbol{o}[t+1]$ and $\boldsymbol{Z}[t]\gets \boldsymbol{Z}[t+1]$, $u\in \mathbb{U}$;}
		\ENDFOR
		\ENDFOR
	\end{algorithmic}
\end{algorithm}

\subsection{Training stage}

In the training stage, we compare MAGRL with the following $3$ benchmarks: 
\begin{itemize}
	\item MAGRL-HoE: This method does not consider HoE at each IoT device. By only considering the battery energy at each IoT device, its reward function is reduced from (\ref{equ: local reward a}) as
	\begin{align}  
		r_{u,0}[t]\!=\!\frac{N_u[t]\sum_{i\in\mathbb{I}_l[t]}(B_i[t \!+\! 1] \!-\! B_i[t])}
		{1 \!+|\mathbb{I}_l[T]|} \! \nonumber \\ +  \xi_2(\! B_u[t \!+\!1]\!-\! B_u^{min}\!).  \label{equ: re-expressed reward} 
	\end{align}
	\item MAGRL-G: This method removes the global training, where the loss function is given in (\ref{equ: Loss local Q}) with $\varepsilon=1$.
	\item MAGRL-HoE-G: This method does not exploit the HoE and the global training, where both (\ref{equ: re-expressed reward}) and (\ref{equ: Loss local Q}) with $\varepsilon=1$ are applied.
\end{itemize}

We consider an area of $400$ m$\times$$400$ m with $4$ UAVs and $6$ IoT devices. For the proposed MAGRL method and the benchmarks, we show the accumulated average reward $r_{ac}=\frac{1}{U}\sum_{t=1}^{T}\sum_{u=1}^{U}r_u[t]$ in Fig. \ref{fig: MAGR performance a}. Fig. \ref{fig: MAGR performance b} shows the variations of $H_{total}$ of all four methods. The convergence of our proposed MAGRL algorithm is observed in Fig. \ref{fig: MAGR performance a}, where the proposed MAGRL method outperforms the other $3$ benchmarks and achieves the highest $r_{ac}$ after convergence. This implies that the global training can learn the potential connections among the states of the UAVs, thus improving the learning ability of multiple agents. From Fig. \ref{fig: MAGR performance b}, it is observed that by considering HoE, $H_{total}$ in our proposed MAGRL method is the lowest among all four methods, which confirms, that the goal of HoE minimization can guide the UAVs' WET to cater to each IoT device's energy requirements.

\subsection{Testing stage}

\begin{figure}[htbp] \label{fig: testing performence}
	\centering 
	\subfigure[UAVs' trajectories.]{ \label{fig: uav-trajectory}
		\begin{minipage}{7.5cm}
			\centering
			\includegraphics[width=7.5cm]{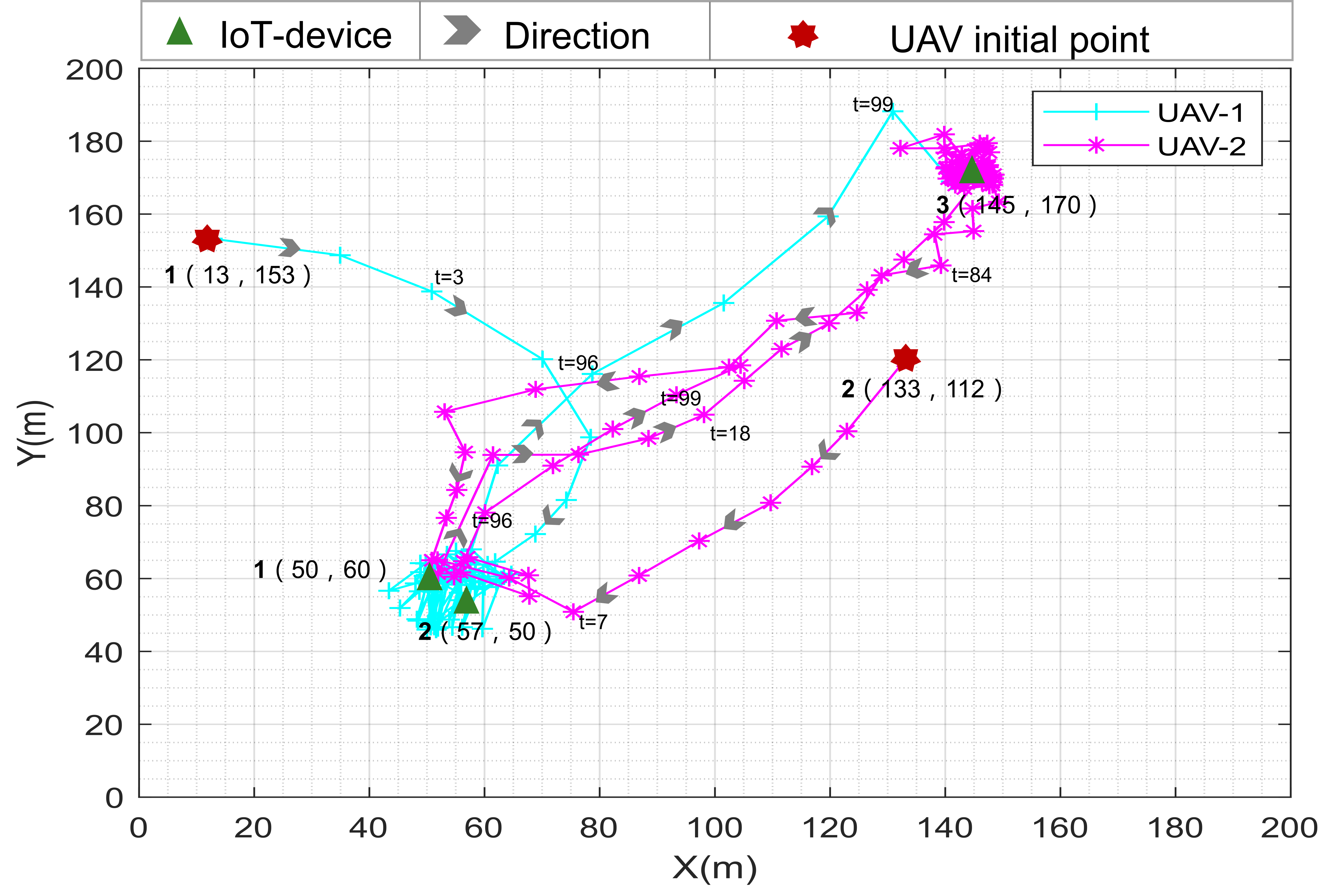}
		\end{minipage}
	}
	\subfigure[IoT devices' battery energy variation.]{ \label{fig: iot-energy}
		\begin{minipage}{7.5cm}
			\centering
			\includegraphics[width=7.5cm]{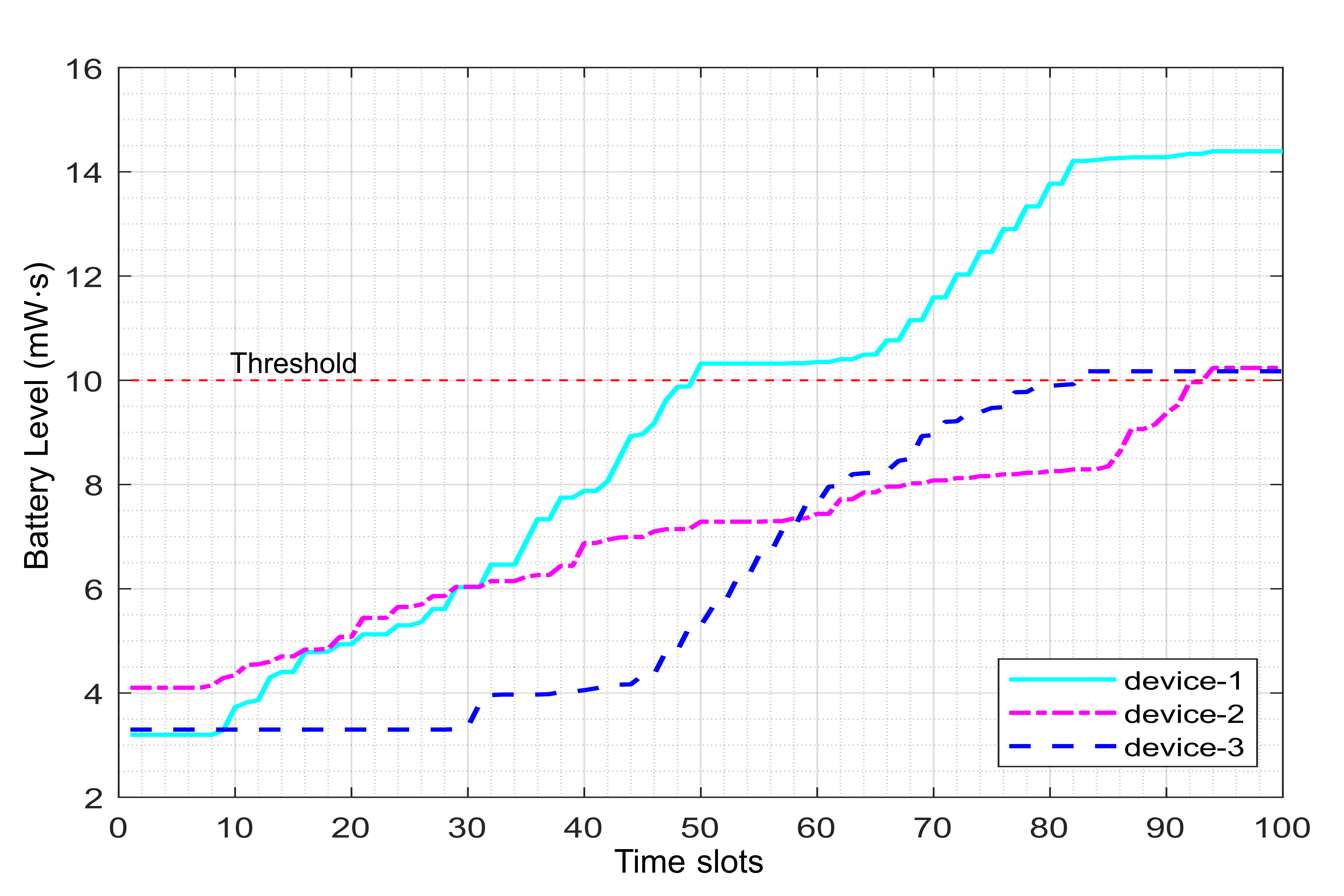}
		\end{minipage}
	}
	\vspace{-0.1in}
	\caption{UAVs' trajectory and the IoT devices' battery energy under the proposed MAGRL method.}
\end{figure}

To show the performance of the UAVs' WET in the testing stage, we illustrate an example, where $2$ UAVs are dispatched to charge $3$ IoT devices in a horizontal area of $200$ m$\times$$200$ m. Each UAV agent applies the trained $Actor_L$ network to determine its actions. Fig. \ref{fig: uav-trajectory} shows the trajectories of the two UAVs, Fig. \ref{fig: iot-energy} shows each of the IoT device's battery variations over time, and Fig. 5 shows the two UAVs' binary WET decisions. It is observed from Fig. \ref{fig: uav-trajectory} and Fig. 5 that although each UAV distributively determines its own trajectory and WET, due to the exploration of their self-attentions in the global training, the two UAVs can automatically serve different IoT devices in a collaborative manner, where UAV-1 transmits energy mainly to the two closely-located IoT devices, while UAV-2 mainly serves the other distantly-located IoT devices. Due to their effective collaboration for WET, it is also observed that each of the IoT device's battery energy achieves the required threshold $B^{thr}$ in Fig. \ref{fig: iot-energy}, and thus the overall HoE in problem (P1) becomes $0$. Moreover, at the end of $T=100$ slots, the remained energy of the two UAVs are $24657.72$ W$\cdot$s and $24538.863$ W$\cdot$s, respectively, where the required $B_u^{min}=20000$ W$\cdot$s in (\ref{equ: constraint c}) are  satisfied for both UAVs.

\vspace{-0.5em}

\begin{figure}[htbp]  \label{fig: uav-et}
	\centering 
	\subfigure[UAV-1's WET decisions over time.]{ \label{fig: uav-et-1}
		\begin{minipage}{7.5cm}
			\centering
			\includegraphics[width=7.5cm]{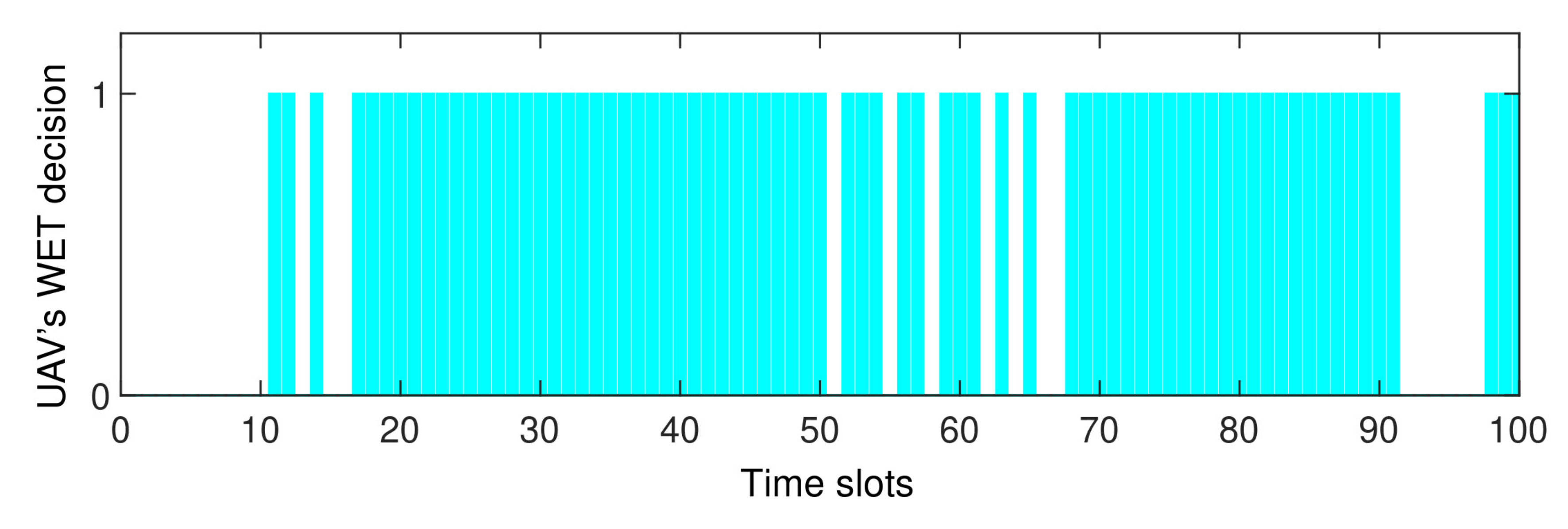}
		\end{minipage}
	}
	\subfigure[UAV-2's WET decisions over time.]{ \label{fig: uav-et-2}
		\begin{minipage}{7.5cm}
			\centering
			\includegraphics[width=7.5cm]{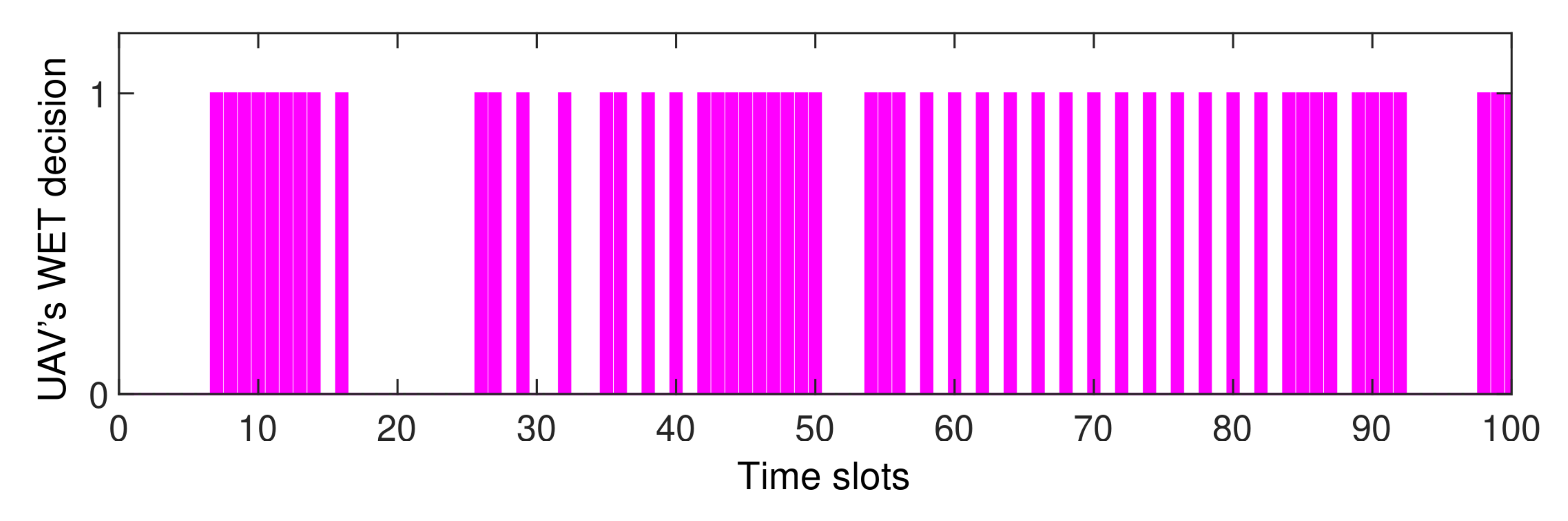}
		\end{minipage}
	}
	\vspace{-0.1in}
	\caption{UAVs' WET decisions under the proposed MAGRL method.}
\end{figure}

\section{Conclusion} \label{section: conclusion}
This paper proposes the novel on-demand WET scheme of multiple UAVs. We propose a new metric of HoE to measure each IoT device's time-varying energy demand based on its required battery energy and the harvested energy from the UAVs. We formulate the HoE minimization problem under practical UAVs' mobility and energy constraints, by optimally determining the UAVs' coupled trajectories and WET decisions over time. Due to the high complexity of this problem, we leverage DRL and propose the MAGRL-based approach, where the UAVs' collocations for WET are exploited by excavating the UAVs' self-attentions in the global training. Through the offline global and local training at the central controller and each UAV, respectively, each UAV can then distinctively determine its own trajectory and WET based on the well-trained local neural networks. Simulation results verify the validity of the proposed HoE metric for guiding the UAVs' on-demand WET, as well as the UAVs' collaborative WET under the proposed MAGRL-based approach.

\section*{Acknowledgement}
This work was supported by the National Natural Science Foundation of China under Grant 62072314.

\newpage


\begin{thebibliography}{11}
\bibliographystyle{IEEEbib}



\bibitem{ref.1} S. Bi, C. K. Ho, and R. Zhang,  ``Wireless powered communication: Opportunities and challenges,'' {\it IEEE Commun. Mag.},  vol. 53, no. 4, pp. 117–125, Apr., 2015.

 
\bibitem{ref.3} B. Clerckx, \emph{et al.}, ``Fundamentals of wireless information and power transfer: From RF energy harvester models to signal and system designs,'' {\it  IEEE J. Sel. Areas Commun.}, vol. 37, no. 1, pp. 4–33, Jan. 2018. 


\bibitem{Yue Ling Che} Y. L. Che, Y. Lai, S. Luo, K. Wu, and L. Duan, ``UAV-aided information and energy transmissions for cognitive and sustainable 5G networks,'' {\it  IEEE Tran. Wireless Commun.}, vol. 20, no. 3, pp.1668-1683, Mar. 2021. 

\bibitem{ref.7}  Z. Yang,W. Xu,M. Shikh-Bahaei, ``Energy efficient UAV communication with energy harvesting,'' {\it  IEEE Trans. Veh. Technol.}, vol. 69, no. 2, pp. 1913-1927, Feb. 2020. 


\bibitem{ref.6} J. Xu, Y. Zeng and R. Zhang, ``UAV-enabled wireless power transfer: trajectory design and energy optimization,'' {\it IEEE Trans. Wireless Commun.}, vol. 17, no. 8, pp. 5092-5106, Aug. 2018.


\bibitem{ref.19} J. Mu and Z. Sun, ``Trajectory design for multi-UAV-aided wireless power transfer toward future wireless systems,'' {\it Sensors}, vol. 22, no. 18, pp. 6859, Aug. 2022.


\bibitem{ref.20} L. Xie, X. Cao, J. Xu and R. Zhang, ``UAV-enabled wireless power transfer: a tutorial overview,''   {\it IEEE Transactions on Green Communications and Networking}, vol. 5, no. 4, pp. 2042-2064, Dec. 2021.




\bibitem {ref.8}  K. Li, W. Ni, E. Tovar and A. Jamalipour, ``On-board deep Q-network for UAV-assisted online power transfer and data collection,'' {\it IEEE Trans. Veh. Technol.}, vol. 68, no. 12, pp. 12215-12226, Dec. 2019.


\bibitem{ref.9} O. S. Oubbati  \emph{et. al.}, ``Synchronizing UAV teams for timely data collection and energy transfer by deep reinforcement learning,'' {\it  IEEE Trans. Veh. Technol.}, vol. 71, no. 6, pp. 6682-6697, Jun. 2022.

%
%
%
%
%
%
%
%
%
%


\bibitem{ref.22} A. Al-Hourani, S. Kandeepan and S. Lardner, ``Optimal LAP altitude for maximum coverage,'' {\it IEEE Wireless Commun. Let.}, vol. 3, no. 6, pp. 569-572, Dec. 2014.  


\bibitem{ref.23} Y. Zeng, J. Xu and R. Zhang, ``Energy minimization for wireless communication with rotary-wing UAV,'' {\it IEEE Trans. Wireless Commun.}, vol. 18, no. 4, pp. 2329-2345, Apr. 2019.


\bibitem{ref.24} P. N. Alevizos and A. Bletsas,  ``Sensitive and nonlinear far-field RF energy harvesting in wireless communications,''  {\it IEEE Trans. Wireless Commun.}, vol. 17, no. 6, pp. 3670-3685, Jun. 2018. 

\bibitem{ref.PowerCast} PowerCast Module. Accessed: Jul. 2020. [Online]. Available: http://www.mouser.com/ds/2/329/P2110B-Datasheet-Rev-3-1091766.pdf


\bibitem{ref.26} U. V. Luxburg, ``A tutorial on spectral clustering,'' {\it Statistics and computing}, vol. 17, no. 3, pp. 395-416, Aug. 2007.



\bibitem{ref.27} A. Vaswani, \emph{et. al.},  ``Attention is all you need,'' Advances in neural information processing systems, 2017.



\bibitem{ref.28} J. Jiang, \emph{et. al.} ``Graph convolutional reinforcement learning,'' in Pro. {\it International Conference on Learning Representations (ICLR)}, Oct. 2018.



\bibitem{ref.30} M. L. Puterman, ``Markov decision processes: Discrete stochastic dynamic programming,'' John Wiley and Sons, 2014.


\bibitem{ref.29} T. Haarnoja, \emph{et. al.} ``Soft Actor-Critic: Off-Policy Maximum Entropy Deep Reinforcement Learning with a Stochastic Actor,'', in Pro. {\it International Conference on Machine Learning (ICML)}, Aug. 2018.


\end{thebibliography}
\end{document}